# COMPARISON OF METHODS FOR CHARACTERIZING SOUND ABSORBING MATERIALS

**Yacoubou SALISSOU and Raymond PANNETON**
GAUS, Department of mechanical engineering, Université de Sherbrooke, Québec, Canada, J1K 2R1
Yacoubou.Salissou@USherbrooke.ca , Raymond.Panneton@USherbrooke.ca

## 1. INTRODUCTION

Open-cell porous materials are widely used as acoustic absorbent media in the transport industry. In order to have a prior knowledge of their acoustic behaviour, different models have been developed [1-4]. Among all these models, the Johnson-Champoux-Allard [2, 4] is widely used because of it simplicity and accuracy. This model uses five parameters (open porosity, static airflow resistivity, tortuosity, viscous characteristic length and thermal characteristic length) which describe the internal structure of porous materials at the macroscopic scale. Different methods have been developed in order to characterize those parameters. Some of them are based on the physical and mathematical definition of the parameters. However, these methods, qualified as direct methods, require dedicated equipments. The second approach is based on the acoustical model from which analytical expressions linking the material parameters to acoustical measurements are derived. The methods using this approach are qualified as analytical inversion or indirect methods. Finally, the last approach is based on an optimisation problem where the material parameters are adjusted in the acoustic model to reproduce acoustical measurements. The methods of this group are qualified as inverse methods.

In this paper, three samples of metal foam are characterized using direct, indirect and inverse methods. The objective is to highlight that the methods may yield large variability in the found parameters. A discussion is given to explain the observed variability and the limitations of the methods.

## 2. CHARACTERIZATION METHODS

Direct measurement of the open porosity is performed using the Archimedes principle following the in-air missing mass method [5]. This method gives a direct measured value of open porosity through measurement of sample weight in vacuum and in air, and predicts the measurement error. In this study, because of the lack of material, the three samples will be put together (as one sample) to satisfied the minimum material volume required by the method. The direct method used to measure the static airflow resistivity is based on the method proposed by Stinson & Daigle [6].

The indirect method (IM) used in this work is based on the technique developed by Panneton & Olny [7,8]. Assuming dynamic density, dynamic bulk modulus, and open porosity known, analytical solutions are used to determine the static airflow resistivity, tortuosity, viscous characteristic dimension (VCD), and thermal characteristic dimension (TCD) of the material.

The inverse method is based on an optimization problem where unknown parameters are adjusted to fit measured acoustic data (ex.: sound absorption coefficient) [9]. This method is first applied to find the tortuosity and the two characteristic lengths by assuming porosity and static resistivity known from direct measurements. Second, the algorithm is applied to find simultaneously the five parameters assuming that none of the parameters are known. The acoustical measurements required for the indirect and inverse methods were obtained using an impedance tube following ASTM E 1050 and ISO 10534-2 standards.

## 3. RESULTS AND DISCUSSION

The dimensions of the samples are summarized in table 1. Direct measured porosity is 0.89±0.03. Using this value, the four other parameters obtained from indirect method are summarized in table 2. In general, the parameters values are in good agreement from one sample to another and the deviation from mean values are acceptable. The deviation of tortuosity and characteristics length may appear a little bit high, but since these parameters are difficult to measure with an accurate precision, these deviations are acceptable. Table 3 shows that the method yields stable results inside the accepted values of porosity, and finally, the static airflow resistivity given by this method fits well with those measured directly (table 4). Using the directly measured porosity and mean value of airflow resistivity, the three-parameter inverse method (3-PIM) yields results summarized in table 5. As the indirect method, the results fit quite well from one sample to another and this method yields stable results inside the accepted values of porosity (table 6) and airflow resistivity (table 7), but seems to depend on frequencies range (table 8). In term of behaviour, the five-parameter inverse method (5-PIM) yields similar results as the three-parameter method (see tables 9 and 10).

Table 11 presents a comparison of the results obtained from the different methods. We observe a significant difference between the porosity value giving by the direct and inverse methods. Moreover, the airflow resistivity obtained by inverse method seems to be over estimated. Finally, the results of the 3-PIM and IM compare well; however they diverge slightly from those giving by the 5-PIM. As 3-PIM and 5-PIM rely only on acoustical measurements, their accuracy exclusively depends on their quality and reproducibility. Usually because of experimental conditions (room temperature, sample fixing, etc), those measurements present a small deviations. The 5-PIM may be strongly vulnerable to those deviations and this may explain its variability. The two other methods (3-PIM and IM) are also affected by acoustical measurement deviations (this may explain their small variability), but less strongly than the 5-PIM since they use some known parameters. Furthermore, because both indirect and inverse methods rely on acoustical measurements, they are less accurate than the direct method. Finally, since both are based on the Johnson-Champoux-Allard model which supposes a motionless frame, they may yield wrong parameters when this condition is not satisfied. These results support the need of: 1) developing non-acoustical and direct methods (mainly for tortuosity and characteristic lengths), or 2) developing more accurate measurements in impedance tube for inverse and indirect characterization purposes.

## ACKNOWLEDGEMENTS

N.S.E.R.C. supported this work.

Table 1: Samples dimensions

| Sample | Thickness (mm) | Diameter (mm) |
|---|---|---|
| 1 | 18.83 | 29.0 |
| 2 | 19.12 | 29.0 |
| 3 | 19.31 | 29.0 |

Table 2: Results from indirect method.
The porosity value was 0.89± 0.03 measured with missing mass method [5] on the three samples at a same time. Uncertainty is predicted by method.

| Sample | Tortuosity | Resistivity (Ns/m4) | VCD (µm) | TCD (µm) |
|---|---|---|---|---|
| 1 | 1.38 | 51 290 | 21.9 | 109.5 |
| 2 | 1.31 | 49 887 | 20.5 | 114.7 |
| 3 | 1.18 | 48 516 | 18.2 | 122.3 |
| **Mean** | **1.29** | **49 898** | **20.2** | **115.5** |

Table 3: Effect of porosity precision on indirect method

| Porosity | Tortuosity | Resistivity (Ns/m$^4$) | VCD (µm) | TCD (µm) |
|---|---|---|---|---|
| 0.88 | 1.28 | 49 840 | 20.2 | 106.0 |
| 0.89 | 1.29 | 49898 | 20.2 | 115.5 |
| 0.90 | 1.30 | 49927 | 20.3 | 132.0 |

Table 4: Static airflow resistivity from direct method

| Sample | 1 | 2 | 3 | Mean |
|---|---|---|---|---|
| Resistivity (Ns/m4) | 51 325 | 50 034 | 48 670 | 50 010 |

Table 5: Results from 3-PIM
Porosity: 0.89± 0.03, resistivity: 50010 Ns/m$^4$, frequency: 800-6000 Hz

| Sample | Tortuosity | VCD (µm) | TCD (µm) |
|---|---|---|---|
| 1 | 1.45 | 24.8 | 119.9 |
| 2 | 1.45 | 24.9 | 122.6 |
| 3 | 1.26 | 21.2 | 162.9 |
| **Mean** | **1.39** | **23.6** | **135.1** |

Table 6: Effect of porosity precision on 3-PIM

| Porosity | Tortuosity | VCD (µm) | TCD (µm) |
|---|---|---|---|
| 0.88 | 1.42 | 24.8 | 122.4 |
| 0.89 | 1.39 | 23.6 | 135.1 |
| 0.90 | 1.34 | 22.1 | 149.0 |

Table 7: Effect of resistivity precision on 3-PIM

| Resistivity (Ns/m$^4$) | Tortuosity | VCD (µm) | TCD (µm) |
|---|---|---|---|
| 49000 | 1.31 | 21.7 | 142.9 |
| 50010 | 1.39 | 23.6 | 135.1 |
| 51000 | 1.43 | 24.5 | 136.0 |

Table 8: Frequency range effect on 3-PIM

| Frequency range (Hz) | Tortuosity | VCD (µm) | TCD (µm) |
|---|---|---|---|
| 800-6000 | 1.39 | 23.6 | 135.1 |
| 800-4000 | 1.37 | 22.6 | 153.5 |
| 300-6000 | 1.66 | 31.3 | 112,2 |
| 300-4000 | - | - | - |
| 1000-3000 | 1.40 | 23.0 | 146.3 |
| 2000-6000 | 1.32 | 21.5 | 125.7 |

Table 9: Results from 5-PIM
Frequency range: 800-6000 Hz

| Sample | Porosity | Resistivity (Ns/m$^4$) | Tortuosity | VCD (µm) | TCD (µm) |
|---|---|---|---|---|---|
| 1 | 0.84 | 55 135 | 1.54 | 28.2 | 85.1 |
| 2 | 0.84 | 54 688 | 1.53 | 28.3 | 85.6 |
| 3 | 0.81 | 53 112 | 1.42 | 28.0 | 85.1 |
| **Mean** | **0.83** | **54 312** | **1.50** | **28.2** | **85.3** |

Table 10: Frequencies range effect on 5-PIM

| Frequency range (Hz) | Porosity | Resistivity (Ns/m$^4$) | Tortuosity | VCD (µm) | TCD (µm) |
|---|---|---|---|---|---|
| 800-6000 | 0.83 | 54 312 | 1.50 | 28.2 | 85.3 |
| 300-6000 | 0.82 | 52 045 | 1.36 | 24.7 | 84.1 |
| 300-4000 | - | - | - | - | - |
| 1500-6000 | 0.89 | 54 050 | 1.47 | 26.5 | 117.8 |
| 2000-6000 | 0.91 | 55 520 | 1.56 | 26.7 | 128.0 |

Table 11: Comparison of results from the different methods

| Method | porosity | Resistivity (Ns/m$^4$) | Tortuosity | VCD (µm) | TCD (µm) |
|---|---|---|---|---|---|
| Direct | 0.89 | 50 010 | - | - | - |
| Indirect | - | 49 898 | 1.29 | 20.2 | 115.5 |
| 3-PIM | - | - | 1.39 | 23.6 | 135.1 |
| 5-PIM | 0.83 | 54 312 | 1.50 | 28.2 | 85.3 |